\documentstyle[aps]{revtex}
\begin{document}
\draft
\date{\today}     
\title{Nuclear Reaction Rates in a Plasma:\\The Effect of Highly Damped
Modes}
\author{Merav Opher$^{1}$\footnote{email: merav@physics.ucla.edu}, Reuven
Opher$^{2}$\footnote{email: opher@orion.iagusp.usp.br}}
\address{${}^{1}$Physics and Astronomy Department\\
T-Plasma Group, 1-130 Knudsen\\
University of California, Los Angeles, CA 90095-1547, U.S.A \\   
${}^{2}$Instituto Astron\^omico e Geof\' \i sico - IAG/USP,\\
Av. Miguel St\'efano, 4200 \\
CEP 04301-904 S\~ao Paulo, S.P., Brazil}
\maketitle
\begin{abstract}
The {\it fluctuation-dissipation theorem} is used to evaluate the screening
factor of nuclear reactions due to the electromagnetic fluctuations in a
plasma. We show that the commonly used Saltpeter factor is obtained
if only fluctuations near the plasma eigenfrequency are assumed to be 
important ($\omega \sim \omega_{pe}\ll T$ ($\hbar=k_{B}=1$)). By taking into 
account all the fluctuations, the highly damped ones, with 
$\omega >\omega_{pe},$ as well as those with 
$\omega\leq\omega_{pe},$  we find that nuclear reaction rates are higher 
than those obtained using the Saltpeter factor, for many interesting plasmas.
\end{abstract}
\pacs{PACS numbers: 52.25.Dg, 95.30.Qd}
\section{Introduction}
It is known that fusion reactions within a plasma  are enhanced. This is
due to the fact that ions in a plasma are not ``naked'', but are
surrounded by electrons in the plasma that form a shielding cloud around 
them. Thus, an ion is not repelled, as is a ``naked'' ion, since the
``dressing'' of electrons shield the repulsive potential of the ion for 
distances greater than the Debye radius, $R_{D}$. The presence of the 
plasma is taken into account, in what is called the 
{\it screening factor}\cite{clay}, in estimating cross sections of nuclear 
reaction rates. In the static case, when the 
velocities of the ions are assumed to be very slow compared to those of 
the screening particles, the screening factor used is the Salpeter factor
\cite{sal} (valid in the weak screening limit, when 
$Z_{1}Z_{2}e^{2} \ll T\, R_{D}(k_B=1)).$ Electromagnetic fluctuations are
present in a plasma, even when in thermal equilibrium. 
From these fluctuations, we evaluate the 
screening of the interacting particles. Their effect is to enhance the 
reaction rates, just as a negative potential reduces the Coulomb repulsion. 
\par
We use the formalism of the fluctuation-dissipation theorem\cite{sit,akh}
to describe the electromagnetic fluctuating fields. Using the 
fluctuation-dissipation theorem, it is possible to study all the 
fluctuations existent in a plasma. That is, it is possible to also include 
the effect of screening modes that are not eigenmodes (i.e., not at the
plasma frequency). These modes are highly damped, but although they have a 
small amplitudes, their overall effect turns out to be great because their 
phase space is very large. In particular, they have non-negligible amplitudes
for $\omega \gg \omega_{pe}$, where $\omega_{pe}$ is the plasma frequency. In
estimating the effect of electromagnetic fluctuations on the screening in the
limit where only frequencies on the order of $\omega_{pe}$ are important
($\omega \sim \omega_{pe} \ll T$ ($\hbar=k_{B}=1)),$ the usual Salpeter factor is
obtained. 
\par
In a non-magnetized plasma, the longitudinal eigenmodes are Langmuir waves 
with $\omega \sim \omega_{pe}.$  For usual laboratory and space plasmas, this 
frequency is much less than $T$ and the assumption that $\omega \ll T$
is reasonable. However, if we are interested in the effect of all the
existent modes in the plasma on the screening, it is important not to be
limited only to modes where $\omega\sim\omega_{pe}.$ 
\par
We show, in section II, that the screening can be described in 
terms of the electromagnetic fluctuations. The change in the reaction rates 
due to the screening is calculated and our screening factor is compared
with the Salpeter factor. 
A discussion of our results is given in section III.
\section{Reaction Rates}
\label{sec:rec}
The expression for the reaction rate between ions 1 and 2 is \cite{clay} 
\begin{equation}
R=\int d{\bf v_{1}} f(v_{1}) \int d{\bf v_{2}} f(v_{2}) 
\mid v_{1} - v_{2} \mid \sigma(v_{1},v_{2})~, 
\end{equation}
where $\sigma$ is the 
cross section and $f(v_{1})$ and $f(v_{2})$ are the thermal
distributions of the particles. The cross section is 
$\sigma(v_{1},v_{2})=(S_{N}(E)/E)P(v_{1},v_{2})$, where 
$P(v_{1},$ $v_{2})$ is the penetration factor,
$E$ is the kinetic energy of the ions at large separations in the 
center of mass reference frame and 
$S_{N}(E)$ is a purely nuclear factor that varies slowly with $E$.
$P(v_{1},$ $v_{2})$ is proportional to 
$exp[( -2 {( 2\mu )}^{1/2})/\hbar )\int_{a}^{r_{0}}$
${[U(r)-U_{0}-E]}^{1/2}dr ]$, 
where $\mu$ is the reduced mass. 
The integration 
is performed from the nuclear radius $a$ to the turning point 
radius $r_{0}$. $U(r)$ is the bare Coulomb interaction 
$Z_{1}Z_{2}e^{2}/r$ and $U_{0},$ the potential energy due to the 
screening. In the case of weak screening, $U_{0}$ is pratically 
independent of $r$ in the integration interval. The reaction rate
$R$ is then found to be increased by a factor 
$F=exp(U_{0}/T)$ \cite{clay}. In the static case, the weak screening
limit, 
the enhancement factor is $exp(Z_{1}Z_{2}e^{2}/(T~R_{D}))$, the 
Salpeter factor.
\par
In a plasma, electromagnetic fluctuations are always present, acting as a
background for the interacting ions. Let us consider
the effect of the fluctuating electric fields. 
$\phi$, the electric potential acting in the 
region from $r_{0}$ to $a$, creates a potential energy
$U_{1}=e(Z_{1}+Z_{2})\phi$. 
The enhancement factor is, then, $F=exp(U_{0}/T)=exp(U_{1}-U_{\infty} /T)$, 
where $U_{\infty}$ is the potential energy of the ions 
when they are at infinity, 
$U_{\infty}=Z_{1}e \phi_{1,\infty}+Z_{2}e\phi_{2,\infty}$. 
Averaging $F$ in time and using a Taylor expansion, we obtain 
terms which are proportional to $\langle \phi^{2} \rangle$, 
$\langle \phi_{1,\infty}^{2} \rangle$ and 
$\langle \phi_{2,\infty}^{2} \rangle$, as well as 
cross terms. Because the fields $\phi$, $\phi_{1,\infty}$ and 
$\phi_{2,\infty}$ are random with respect to one another, the 
only terms that remain to first order are $\langle \phi^{2} \rangle$, 
$\langle \phi^{2}_{1,\infty} \rangle$ and $\langle \phi^{2}_{2,\infty}
\rangle$. 
Assuming that they are equal to one another,   
\begin{equation}
{\it{F}}=1+Z_{1}Z_{2}e^{2}\langle \phi^{2} \rangle \beta^{2} +...= 
exp \left ( Z_{1}Z_{2}e^{2}\langle \phi^{2} \rangle 
\beta^{2} \right ) ~,
\end{equation}
where $\beta=1/T$. 
\par
$\phi$ is assumed to be changing slowly 
in the integrand of the penetration factor in the 
region from $r_{0}$ to $a$. Therefore, we consider 
fields with frequencies $\omega < \omega_{max}=2\pi r_{0}/v$ and
wavenumbers 
$k <k_{max}=2\pi/r_{0}$, 
where $v$ is the velocity of the reduced mass in the center of mass
system. 
We find that $\langle \phi^{2} \rangle$ is relatively insenstive to the 
exact upper limits of the integrations, $\omega_{max}$ and $k_{max}$. 
\par
$\langle \phi^{2} \rangle$ is related to the 
fluctuations of the longitudinal electric field in a plasma: 
${\langle \phi^{2} \rangle}_{k}={\langle E^{2} \rangle}_{k}/k^{2}$. 
Therefore, 
\begin{equation}
\langle \phi^{2} \rangle=\int \frac{d{\bf k}}{(2\pi)^{3}}
{\langle \phi^{2} \rangle}_{k}=\int \frac{d{\bf k}}{(2\pi)^{3}} 
\int d\omega \frac{{\langle E^{2} \rangle}_{k\omega}}{k^{2}}~.
\label{phi}
\end{equation} 
>From the fluctuation-dissipation theorem\cite{sit,akh}, the expression 
for the intensity of the fluctuations of the longitudinal electric field
is
\begin{equation} 
\frac{{\langle E^{2} \rangle}_{k \omega}}{8\pi}= 
\frac{1}{exp({\omega/T})-1} 
\frac{Im \varepsilon_{l}}{{\mid \varepsilon_{l} \mid}^{2}}~,
\label{elf}
\end{equation} 
where $\varepsilon_{l}$ is the longitudinal dielectric permittivity of the
plasma. We note that this is an exact quantum mechanical relation. Given a 
dielectric permittivity that describes the plasma, it is possible, from 
Eq. (4), to obtain the fluctuations in the longitudinal field due to the
eigenmodes (i.e., the electrostatic Langmuir 
waves), as well as to all the other fluctuations existent in the plasma. 
\par
We show below that the overall effect of fluctuations with
$\omega >\omega_{pe}$ is to substantially increase 
the reaction rates. Although the amplitudes of these fluctuations are small,
they are not negligible. Due to the very large range of their frequencies, 
their overall effect is to increase the reaction rates appreciably. 
Because the reaction rates are exponentially dependent
on the screening potential, its increase is appreciable
for reactions with large values of $Z_{1}Z_{2}$ such as $p-{}^{7}Be$ and 
$p-{}^{14}N$. In the limit where only fluctuations with frequencies on the 
order of the eigenfrequency $\omega_{pe}\;(\ll T)$ are important, Eq. (4)
can be easily intergrated over frequency with the aid of the Kramers-Kronig
relations, obtaining
\begin{equation}
{\langle E^{2} \rangle}_{k}=\frac{T}{2} \left ( 1-
\frac{1}{\varepsilon_{l}(0,k)} \right )~.
\end{equation}
This is the expression which is generally found in text books
\cite{sit,akh}. For
an electron plasma in the collisionless case, the longitudinal dielectric
permittivity is $\varepsilon_{l}(\omega,{\bf
k})=1+\frac{k^{2}_{D}}{k^2}\left ( 1-\phi(z)+i\sqrt{\pi}ze^{-z^{2}} \right
)$, where $z=\sqrt{3/2}(\omega/kv_{T})$ and $\phi(z)=2~z~e^{-z^{2}}
\int_{0}^{z}
e^{x^{2}} dx$. With this dielectric permittivity,
$\langle \phi^{2} \rangle =T/R_{D}$ [Eq.(3)] and the enhancement factor F, 
in Eq.(2), is the Salpeter enhancement factor, 
\begin{equation}
{\it F_{SAL}}=exp \left ( \frac{Z_{1}Z_{2}e^{2}}{T}\frac{1}{R_{D}} \right
) ~.
\end{equation} 
\par
We use the model described in detail 
in Opher \& Opher \cite{op1,op2,op3}, which includes thermal and collisional
effects, to obtain the longitudinal dielectric permittivity. It uses the
Vlasov equation in first order with the BGK (Bhatnagar-Gross-Krook) 
collision term, which is a model equation 
of the Boltzmann collision term \cite{cle}. The inclusion of collisions 
changes the results very little and the effect on 
$\varepsilon_{l}(\omega,{\bf k})$ for the plasmas studied, is almost
negligible. For example, in the case analysed 
in Opher and Opher \cite{op3}, where
$T=10^{5}\;\rm{K}$ and $n=10^{10}\;\rm{cm^{-3}}$, the difference in 
the correlation energy, with or without collisions, is less than
$10^{-6}$.  
\par
The dielectric permittivity that we therefore use, is almost
identical to the collisionless dielectric permittivity given after Eq. (5)
in the text (the only difference is the inclusion of other shielding
species besides the electron). From this model, the longitudinal 
dielectric permittivity for an isotropic plasma is found to be  
\begin{equation}
\varepsilon_{l}(\omega,{\bf k}) = 1 + \sum_{\alpha} 
\frac{{\omega_{p\alpha}}^{2}}{k^{2}{v_{\alpha}}^{2}}
\frac{1+\frac{(\omega+i\eta)}{\sqrt{2}kv_{\alpha}}Z \left ( 
\frac{\omega+i\eta_{\alpha}}{\sqrt{2}kv_{\alpha}} \right )}
{1+\frac{i\eta}{\sqrt{2}kv_{\alpha}}Z \left (
\frac{\omega+i\eta_{\alpha}}{\sqrt{2}kv_{\alpha}} \right )}~,
\label{el}
\end{equation}
where $\alpha$ is the label of the species of the particles, $v_{\alpha}$ is
the thermal velocity of the species and $Z(z)$, the Fried and Conte 
function.
By using Eq. (\ref{el}) in Eq. (\ref{phi}) and Eq. (\ref{elf}), 
we obtain $\langle \phi^{2} \rangle.$ The enhancement factor is obtained 
from Eq. (2). We note that using a relation which is similar to Eq. (2) for 
the transverse field $\langle B^2\rangle_{k,\omega}$ in conjunction with the
transverse dielectric permittivity instead of Eq. (7), the correct black body
spectrum is obtained [5,6].
\par
We estimate the enhancement factor ${\it F}$ for the reaction rates 
$p-p$, ${}^3He-{}^3He$ and $p-{}^{14}N$ for an electron-proton plasma at
temperatures $T=10^{7}-10^{8}\;\rm{K}$ and densities ranging from
$10^{23}-10^{28}\;\rm{cm^{-3}},$ choosing the maximum density for each
temperature
so that the plasma parameter is less than $g=1/n{\lambda_{D}^3}
\sim 0.8$. When $g$ approaches such high values, non-linear effects begin
to be important. (Note that $\lambda_{D}$ has an extra factor $\sqrt{2}$
due to the inclusion of protons.)
\par
The Gamow energy, the most effective energy for thermonuclear
reactions, is 
$E_{G}=1.22 ((Z_{1}Z_{2})^{2}AT_{6}^{2})^{1/3}\;\rm{keV}$, 
where $A$ is the reduced atomic weight and $T_{6}=T/10^{6}$. Therefore,
for each of the temperatures that we choose, the Gamow energy changes, with
a consequent change in the limits of the integration, $k_{max}$ and
$\omega_{max}$. As a particular case, we assume the center 
of mass energy to be zero.
$(F-F_{SAL})/F_{SAL} (\%)\;{\rm{vs}}\; g$, where $F_{SAL}$ is the Salpeter 
enhancement factor and $g$ is the plasma parameter, is plotted for
$T=10^{7}\;\rm{K}$ in Fig. 1. 
In Fig. 2, we plot $(F-F_{SAL})/F_{SAL} (\%)$ vs $g$ for $T=10^{8}\;\rm{K}.$ 
We also estimate the enhancement factor $F$ for the reaction $D-D$ in an
electron-deuteron plasma for temperatures $T=10^{7}-10^{8}\;\rm{K}$ and
densities $10^{23}-10^{28}\;\rm{cm^{-3}}.$ Again, the maximum density
for each temperature is chosen to be $g \sim 0.8$. 
$(F-F_{SAL})/F_{SAL} (\%)$ vs $g$ is shown in Fig. 3.
We see that the enhancement factor $F$ 
is larger than the Salpeter factor. 
\par  
In order to see how the fluctuating potential is affected by these modes,
we plot the potential ${\langle
{\phi}^{2}\rangle}_{\omega,k}$ vs $\omega/\omega_{pe}$ for the reaction
$p-{}^{14}N$, with $T=10^{7}\;\rm{K}$ and $n=10^{24}\;\rm{cm^{-3}}$ (in this
case, $g=0.3$) in Fig. 4. The solid curve is plotted for $k=100~k_{D}$, the 
dashed curve
for $k=10~k_{D}$ and the dotted curve for $k=k_{D}$. We normalize the
curves and plot ${\langle {\phi}^{2} \rangle}_{\omega,k}$ divided by 
${\langle {\phi}^{2}\rangle}_{k,\omega=0}$. When 
${\langle {\phi}^{2} \rangle}_{\omega,k}$ is integrated for all wavenumbers
$k$, but only for frequencies up to $\sim \omega_{pe}$ ($\ll
T$), we obtain the Salpeter potential. We see that for $k=k_{D}$, the
potential $\langle \phi^2 \rangle_{\omega,k}$ has non-negligible values 
only at low frequencies. However, for $k=10~k_{D}$
and $k=100~k_{D}$, the potential $\phi_{k,\omega}$ spreads out to high
frequencies. For these high values of $k,$ instead of dropping abruptly at
$\omega \sim \omega_{pe},$ the curves decrease slowly with frequency. There 
is thus a substantial
contribution for $\omega >\omega_{pe}$. Thus for large wavenumbers,
(i.e., $k>k_{D}$) fluctuations make a substantial contribution at
frequencies higher than the plasma frequency. 
For example, for $k=10\;k_D$ and
$\omega =20\;\omega _{pe},$ $\langle\phi^2\rangle_{\omega, k}/\langle\phi^2
\rangle_{\omega=0,k}$ is not zero, but 0.002. Similarly, for $K=100\;k_D$
and $\omega = 20\;\omega_{pe},$ $\langle\phi^2\rangle_{\omega, k}/\langle
\phi^2\rangle_{\omega=0,k}$ is 0.01.
\section{Conclusions and Discussion}
\label{sec:co}
By using the {\it fluctuation-dissipation theorem} to estimate the 
screening of the reacting ions due to the electromagnetic field fluctuations
in plasma, we calculate the enhancement factor for the reaction rates. The 
enhancement is found to be greater than that for
the adiabatic case ($\omega \ll T$), where the Salpeter 
factor is valid. It is important to note that this effect is not due to a 
specific choice of the dielectric permittivity. The model that we use
includes the standard dielectric permitivity and gives almost the same
results as does a collisionless thermal description. A higher enhancement
factor is obtained because we include the high frequencies and do not 
assume that $\omega \leq\omega_{pe}\ll T$. Using this method in a previous
study to derive the transverse magnetic fluctuations, the standard black 
body spectrum was obtained.
For the plasmas studied here, electron-proton plasmas with temperatures 
$10^{7}-10^{8}\;\rm{K}$ and densities 1$0^{23}-10^{28}\;\rm{cm^{-3}},$ 
the reaction rates for $p-p$, ${}^3He-^3He$ and $p-{}^{14}N$ are increased 
up to $8-13\%$ in the regime $g<1$. The reaction $D-D$ for an
electron-deuteron plasma is increased by $1-2\%$. We find that these
fluctuations have small, but not negligible amplitudes, compared to those 
at low frequencies. They exist, however, in a much larger range of phase 
space, up to very high frequencies. Their overall effect is to increase the
screening potential. Because the reaction rates are exponentially dependent
on the screening potential, this increase is especially strong for reactions
with relatively large $Z_{1}Z_{2}$, such as $p-{}^{7}Be$ and $p-{}^{14}N$. 
When summing the fluctuations only up to the plasma eigenfrequency
$\omega \sim \omega_{pe},$ the Salpeter enhancement factor is obtained. We
show here that the fluctuating potential has a large contribution from highly 
damped modes with frequencies $\omega > \omega_{pe}$.
\par
There has been some controversy about the existence of a screening factor, 
which is different from the Salpeter factor, due to dynamic 
screening \cite{dyn}. The effect described here is due to quasi-fluctuations
that are heavily damped and that always exist in a plasma . They alter the
Salpeter screening factor, although not necessarily in the same way as does
dynamic screening. 
\par
This work shows that by taking into account all the modes, including those
which are heavily damped, with frequencies higher than the plasma frequency,
the screening factor of the nuclear reaction rates is altered and is
larger than the Salpeter factor. The consequences of this work are left for
future studies.
\par
R.O.  would like to thank the Brazilian agency CNPq for 
partial support. M.O would like to thank the U.S. Department of Energy for support (grant
De-FG03-93ER54224).

\begin{figure}[htbp]
\caption{The expression $(F-F_{SAL})/F_{SAL}~(\%)$ vs $g$ for an electron-proton
plasma at a temperature $T=10^{7}\;\rm{K}$. $F$ is the enhancing factor, 
which is compared to the Salpeter enhancing factor $F_{SAL}.$  $g$ is the
plasma parameter $1/n\lambda_{D}^{3}$. The solid curve is the $p-p$
reaction, the dashed curve is the ${}^3He-{}^3He$ reaction, and the dotted
curve, the $p-{}^{14}N$ reaction.} 
\label{fig1}
\end{figure}
\begin{figure}[htbp]
\caption{The expression $(F-F_{SAL})/F_{SAL}~(\%)$ vs $g$ for an electron-proton
plasma at a temperature $T=10^{8}\;\rm{K}$. $F$ is the enhancing
factor, which is compared to the Salpeter enhancing factor $F_{SAL}.$  $g$ 
is the plasma parameter $1/n\lambda_{D}^{3}$. The solid curve is the $p-p$
reaction, the dashed curve is the ${}^3He-{}^3He$ reaction, and the dotted
curve, the $p-{}^{14}N$ 
reaction.} 
\label{fig2}
\end{figure}
\begin{figure}[htbp]
\caption{The expression $(F-F_{SAL})/F_{SAL}~(\%)$ vs $g$ for an
electron-deuteron plasma. $F$ is the enhancing factor,which is compared to 
the Salpeter enhancing factor $F_{SAL}.$  $g$ is the plasma parameter
$1/n\lambda_{D}^{3}$. The solid curve is the reaction $D-D$ for
$T=10^{7}\;\rm{K}$ and the dashed curve, the reaction $D-D$ for 
$T=10^{8}\;\rm{K}$.} 
\label{fig3}
\end{figure}
\begin{figure}[htbp]
\caption{The potential
${\langle {\phi}^{2} \rangle}_{\omega,k}/{\langle {\phi}^{2}
\rangle}_{\omega=0,k}$ vs $\omega/\omega_{pe}$ for the reaction
$p-{}^{14}N$ in an electron-proton plasma at $T=10^{7}\;\rm{K}$  and
$n=10^{24}\;\rm{cm^{-3}}$. The solid curve is for $k=100~k_{D}$, the dashed
curve is for $k=10~k_{D}$, and the dotted curve, for $k=k_{D}$.} 
\label{fig4}
\end{figure}
\end{document}